\documentclass[dvips]{article}

\usepackage{icrc2011}

\title{TeV gamma-ray survey of the northern sky using the ARGO-YBJ experiment}

\newcommand{\etal}{\MakeLowercase{\textit{et al. }}} 
\shorttitle{Z. Cao \etal  TeV $\gamma$-ray Sky survey using ARGO-YBJ}

\authors{Z. Cao$^{1}$, S.Z. Chen$^{1}$, on behalf of the ARGO-YBJ collaboration}
\afiliations{$^1$Particle \& Astrophysics Center,
   Institute of High Energy Physics, CAS}
\email{chensz@ihep.ac.cn}

\abstract{The ARGO-YBJ experiment is an extensive air shower array with full coverage RPC detectors located at Yangbajing
(4300 m asl, Tibet, China). It is operated with high duty cycle ($>$ 86\%) and a large field of view ($\sim$ 2sr). It continuously monitors the entire overhead sky at $\gamma$-ray energies above 0.1 TeV. In the talk, we will present the result of the northern sky survey (between declinations of -10$^{\circ}$ and 70$^{\circ}$) from an analysis of $~$4 years of the ARGO-YBJ data (between July 2006 and February 2011). There are four known TeV sources  observed with significance greater than 5 S.D.. The significance from Crab Nebula is more than 16 S.D.. 90\% confidence level upper limits to the flux from all directions in the sky are also presented, which vary from 0.09 to 0.53 Crab unit for Crab-like point sources. }


\begin{document}
\maketitle

\section{Introduction}
In the last decade, great advances have been made in ground-based $\gamma$-ray astronomy and more than 100 very high energy (VHE) $\gamma$-ray sources have been observed. A new window using VHE $\gamma$-rays is progressively established to probe the non-thermal universe and the extreme physics processes in the astrophysical sources. The VHE $\gamma$-rays are the emissions of the relativistic particles, which are accelerated by the astrophysical shocks that are widely believed to exist in the pulsar wind nebulae (PWN), shell type supernova remnants (SNR), active galactic nuclei (AGN) and so on. These shocks may accelerate protons or electrons. Relativistic electrons can scatter the low energy photons to VHE band, i.e. Inverse Compton (IC) emission, while relativistic protons would lead to  hadronic cascades and VHE $\gamma$-rays are generated by the decay of secondary $\pi^{0}$ mesons. Hence, VHE $\gamma$-ray observations are also important in understanding the origin and acceleration of cosmic rays.

Up to now, the number of  VHE  $\gamma$-ray sources is 107, which includes 61 Galactic sources and 46 extragalactic sources \cite{VHEweb}. Most of the identified Galactic sources belong to  PWN,  SNR, and binary systems, however, about one third of them are still unidentified. The extragalactic sources are mainly composed of blazars, including  BL Lac type objects and flat spectrum radio quasars. 4 radio galaxies and 2 starburst galaxies are also observed.  Due to the absorption by the extragalactic background light (EBL), which causes a substantial reduction of the flux, VHE $\gamma$-ray observations are limited to nearby sources. The most distant source to date is 3C 279 with z=0.536 \cite{albert08}.

Recent advances in VHE $\gamma$-ray observation are mainly attributed to the successful operation of image atmospheric Cherenkov telescopes (IACTs),
such as H.E.S.S., MAGIC, VERITAS and CANGAROO, which made most of the discoveries when searching for counterparts to sources discovered at lower energies.
To achieve an overall view of the universe in VHE $\gamma$-ray band, an unbiased sky survey is needed just like what has been done by Fermi and
its predecessor EGRET at GeV $\gamma$-ray band, which detected 1451 and 271 objects, including 630 and 170 unidentified, respectively \cite{abdo10,hartman99}.
The H.E.S.S. collaboration has made great progress in surveying the inner part of the Galactic plane and discovered  14
$\gamma$-ray sources \cite{aharon06}, however, due to the limitation of small field of view (FOV) and low duty cycle,
for IACTs is difficult to perform a comprehensive sky survey. As a result, although with sensitivity lower than that of IACT, an
extensive air shower (EAS) array, such as the Tibet AS$\gamma$, Milagro and ARGO-YBJ experiments, is the only  choice to perform
a continuous sky survey at VHE band. To date, several surveys of the VHE sky have been performed by  AIROBICC in 2002 \cite{aharon02}, Milagro in
2004 \cite{atkins04} and Tibet AS$\gamma$ in 2005 \cite{amenom05}. The latter two surveys resulted in the successful observation of
$\gamma$-ray emission from Crab Nebula and Mrk421. The best upper limits at energies above 1 TeV are around 275$\sim$600 mcrab  achieved by the Milagro experiment. In 2007, Milagro updated its survey of the Galactic plane and 3 new extended sources were discovered \cite{abdo07a}.
The ARGO-YBJ detector has a better sensitivity than any previous EAS array at 1 TeV and
is continuously monitoring the northern sky with declinations from -10$^{\circ}$ to 70$^{\circ}$.
This paper presents the TeV $\gamma$-ray survey of the northern sky using the 4-year data of ARGO-YBJ.

\section{The ARGO-YBJ experiment}
The ARGO-YBJ experiment, located in Tibet, China, at an altitude of 4300 m a.s.l., is the result of a collaboration among Chinese and Italian institutions and is designed for VHE $\gamma$-ray astronomy and cosmic ray observations. The detector consists of a single layer of Resistive Plate Chambers (RPCs).
130 clusters (a cluster is composed of 12 RPCs and each RPC is composed of 10
logical pads) are installed to form a carpet
of about 5600 m$^{2}$ with an active area of $\sim$93\%. This central
carpet is surrounded by 23 additional clusters (``guard ring'') to
improve the reconstruction of the shower core location. The total area of the
array is  110 m $\times$ 100 m. More details about the detector and
the RPC performance can be found in \cite{aielli06,aielli09b}. The high granularity of the apparatus permits a detailed space-time reconstruction of the shower profile and therefore of the incident direction of the primary particle. The arrival times of the particles are measured by Time to Digital Converters (TDCs) with a resolution of 1.8 ns. This results in an angular resolution of 1.6 degrees for showers with $N_{pad}\simeq20$ and better for increasing $N_{pad}$, tending to 0.3 degrees for showers with $N_{pad}>1000$ \cite{iuppa09}. In order to calibrate the 18,360 TDC channels, an off-line method \cite{he07} has been developed using cosmic ray showers. The calibration precision is 0.4 ns and the procedure is applied every month \cite{aielli09}.

The central 130 clusters began taking data in July 2006, and the ``guard ring'' was merged into the DAQ stream in November 2007. The trigger rate is 3.5 kHz with a dead time of 4\% and the average duty cycle is higher than $86\%$.

\section{Data analysis}

The ARGO-YBJ data used in this analysis were collected from 2006 July to 2011 February. The total effective observation time is 1265 days. To achieve a better angular resolution, the event selections for 6  groups with different $N_{pad}$ used in \cite{barto11} are applied here and only events with zenith angle less than 50 degrees are used. The total number of events after being filtered used in this work is 1.7$\times$10$^{11}$. No $\gamma$/proton discrimination is made
in this analysis.
 In order to obtain a sky map, the celestial coordinates (right ascension and declination) are divided into a grid of $0.1^{\circ}\times0.1^{\circ}$ bins and filled with detected events according to their reconstructed directions. In order to extract an excess of $\gamma$-rays from each bin, the so-called ``direct integral method'' \cite{fleysher04} is adopted to estimate the number of cosmic ray background events in the bin. To remove the affection of large scale anisotropy, a correction has been applied which can be found in \cite{barto11}. To take into account the PSF of the ARGO-YBJ detector, the events in a circular area centered on the bin with an angular radius of $\psi_{70}$ are summed together. The Li-Ma formula \cite{li83} is used to estimate the significance.

\section{Results}
\begin{figure}[!t]
  \vspace{5mm}
  \centering
  \includegraphics[width=3.in,height=2.in]{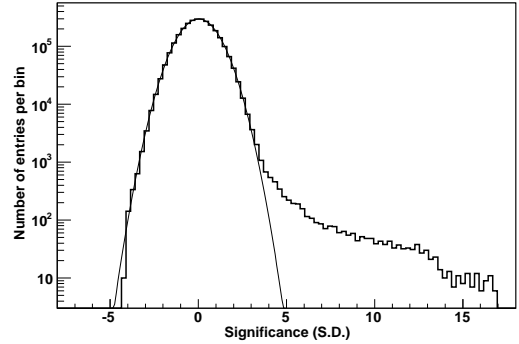}
  \caption{The thick solid line is the distribution of significances from all directions on the northern sky map with declinations from -10$^{\circ}$ to 70$^{\circ}$ . The thin solid line represents the best Gaussian fit to the distribution of significances excluding those cells that have a distance to the Crab, Mrk421, MGRO J1908+06 or MGRO J2031+41 shorter than 4 degree.}
  \label{fig1}
 \end{figure}
The significance distribution from all directions with declinations from -10$^{\circ}$ to 70$^{\circ}$ is shown in Figure 1.  There are many excess points  with significance greater than 5 s.d., which can be attributed to the four known VHE $\gamma$-ray sources, i.e.  Crab, Mrk421, MGRO J1908+06 and MGRO J2031+41 (or TeV J2032+4130). The distribution is consistent with the expectation from random background fluctuations after removing the regions  within 4 degree from the four known sources. A Gaussian fit to this distribution has a mean of 6.5$\times$10$^{-3}$ and a $\sigma$=0.99.

Figure 2 shows the significance map of the northern sky for the ARGO-YBJ data set. There are four sources clearly visible in the map.
Table 1 lists the location of all regions with significance greater than 5 S.D.. For each independent region, only the bin with the largest significance is presented in Table 1. The position is consistent with the ARGO-YBJ pointing error 0.2$^{\circ}$.

\begin{table}[t]
\begin{center}
\begin{tabular}{cc|c|c}
\hline
R.A. (deg)& Dec. (deg) &  S (S.D.) & Source   \\
\hline
83.75  & 22.15 & 17.9  & Crab Nebula \\
166.25  & 38.25   & 13.2 & Mrk 421 \\
286.85  & 6.55 & 5.8  &MGRO J1908+06\\
307.85    & 41.95   & 6.3    &MGRO J2031+41\\
& & & TeV J2032+4130\\
\hline
\end{tabular}
\caption{Location of all regions with an excess greater than 5 S.D.}\label{table1}
\end{center}
\end{table}

  \begin{figure}[!t]
  \centering
  \includegraphics[width=3.in,height=2.in]{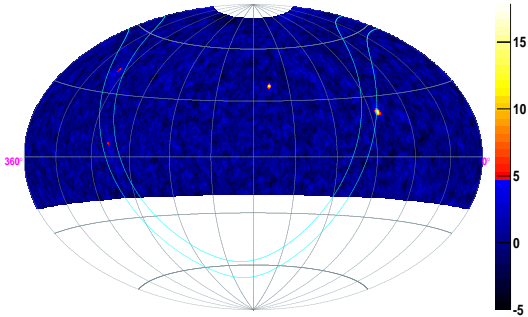}
  \caption{Northern sky as seen by the ARGO-YBJ experiment in TeV $\gamma$-rays. The significance of the excesses in standard deviations is shown by the color scale on the right side. The two light blue lines indicate the Galactic latitudes $\pm$5$^{\circ}$.    }
  \label{fig2}
 \end{figure}

Recently, the Milagro experiment observed 14 of the 34  selected Fermi sources  at a pre-trial
significance of 3 S.D. or more at the representative energy of 35 TeV \cite{abdo09c}. A similar result is also obtained by the AS$\gamma$ experiment \cite{ameno10}. As an  experiment with lower energy threshold than Milagro and AS$\gamma$, ARGO-YBJ observation on these sources can provide more information. Table 2 lists the ARGO-YBJ results for the 15 candidate sources, where the Milagro result for LAT PSR J2238+59c is given in \cite{ameno10}. 14 out of 15 sources are observed with significance greater than 0 S.D.. 9 out of 15 sources  are observed with significance greater than 1.5 S.D., against 1 expected source from the normal Gaussian. The
chance probability from Poisson statistics would be estimated as $10^{-6}$.
It  clearly shows that these sources have
statistically significant correlations with the TeV $\gamma$-ray excesses.
\begin{table*}
\begin{center}
\caption{ARGO-YBJ result for sources observed by Milagro with significance greater 3 S.D.. }\label{table2}
\begin{tabular}{cccccccc}
\hline
Name  & Type & R.A.  & Dec.   &  Milagro  & ARGO-YBJ  & AS$\gamma$ & TeV  \\
 (0FGL) &   &   (deg)&  (deg) &   (S.D.)&  (S.D.) &   (S.D.) &  Assoc.\\
\hline
J0534.6+2201 & PSR & 83.65 & 22.02 & 17.2 &17.8 &6.9   & Crab \\
J0617.4+2234 & SNR & 94.36 & 22.57 & 3.0  &0.7  &0.2   & IC443 \\
J0631.8+1034 & PSR & 97.95 & 10.57 & 3.7  &0.7  &0.3  & \\
J0634.0+1745 & PSR & 98.50 & 17.76 & 3.5  &1.8  &2.2  & Geminga, MGRO C3 \\
J1844.1-0335 &     & 281.04 &-3.59  & 4.3 &2.1  &  & \\
J1900.0+0356 &     & 285.01 & 3.95 & 3.6  &1.6  &1.0   & \\
J1907.5+0602 & PSR & 286.89 & 6.03 & 7.4  &5.5  &2.4  & MGRO J1908+06 \\
J1923.0+1411 & SNR & 290.77 & 14.19 & 3.4 &2.0  &-0.3   & H.E.S.S. J1923+141 \\
J1954.4+2838 & SNR & 298.61 & 28.65 & 4.3 &0.7 &0.6   & \\
J1958.1+2848 & PSR & 299.53 & 28.80 & 4.0 &1.7  &0.1   & \\
J2020.8+3649 & PSR & 305.22 & 36.83 & 12.4 &0.9 &2.2    & MGRO J2019+37 \\
J2021.5+4026 & PSR & 305.40 & 40.44 & 4.2  &3.9 &2.2  & \\
J2032.2+4122 & PSR & 308.06 & 41.38 & 7.6  &5.8 &2.4  & MGRO J2031+41 \\
J2229.0+6114 & PSR & 337.26 & 61.24 & 6.6  &-0.8 &   & MGRO C4 \\
\hline
LAT PSR J2238+59c&PSR & 339.56 & 59.08  & 4.7 &1.2 &2.5  & \\
\hline
\end{tabular}
\end{center}
\end{table*}

To estimate the energy response and the sensitivity of the ARGO-YBJ detector, we simulate different sources in the sky with different declinations. Each source is traced by means of a complete transit, i.e., 24 hr of observation. The simulated events are sampled in the energy range from 10 GeV to 100 TeV, then the full Monte Carlo simulation of the detector is performed. Figure 3 shows the median energy of $\gamma$-rays that trigger ARGO-YBJ and satisfy the event selections as a function of the declination of the source for several spectral indices. When the index is -2.6, which is similar to that of the Crab Nebula, the median energy varies from 0.56 TeV at Dec.=30$^{\circ}$ (the latitude of ARGO-YBJ) to 2.2 TeV at Dec.=-10$^{\circ}$ and Dec.=70$^{\circ}$, while for sources with a hard spectral index -2.0, the corresponding variation of median energy is from 1.1 TeV to 4.5 TeV, and the median energy  varies from  0.36 TeV to 1.1 TeV for sources with a soft spectral index -3.0.

\begin{figure}[!t]
  \centering
  \includegraphics[width=3.in,height=2.3in]{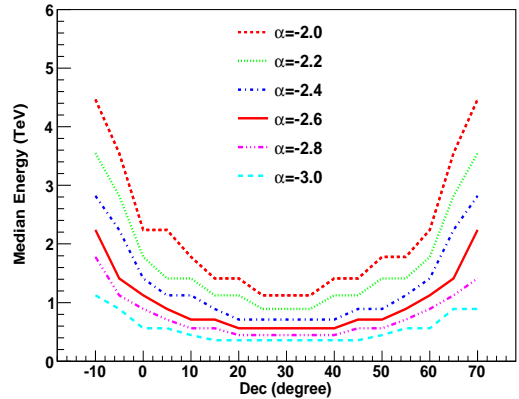}
  \caption{The median energy of $\gamma$-ray events that trigger ARGO-YBJ and satisfy the event selections as a function of source declination. Different lines indicate different spectral indices from -2.0 to -3.0.}
  \label{fig3}
 \end{figure}

Except the four sources listed in Table 1, the significances from all directions are not high enough to claim any signal, therefore we set a  90\% confidence level (C.L.) upper limit to the flux from all directions in the sky. To obtain a good data sample with the same detector configuration, only events collected with 153 clusters  from November 2007 to February 2011, are used to estimate the upper limit.
Firstly, the statistical method given by Helene \cite{helen83} is used to calculate the upper limit on the number of signal events at 90\% C.L. from each bin. Then the number of events is transformed  into flux using the full MC simulation taking into account the daily path of bins in the sky.
Figure 4 gives the average upper limits to the flux of $\gamma$-rays with energies above 100 GeV along the right ascension direction as a function of declination, which vary between  0.09 and 0.53 Crab Unit, i.e. 9.90$\times$10$^{-10}$ cm$^{-2}$ s$^{-1}$, the best ones in the world up to now.  To investigate the flux upper limit for a source with a different spectrum, the upper limits derived assuming a spectral index -2.0 and -3.0 are also presented in Figure 4. The best averaged upper limit for sources with index -2.0 is 3\% Crab Unit, while the similar  value for sources with index -3.0 is 20\% Crab Unit. The upper limits, therefore, depend on the source spectrum.

\begin{figure}[!t]
  \centering
  \includegraphics[width=3.in,height=2.3in]{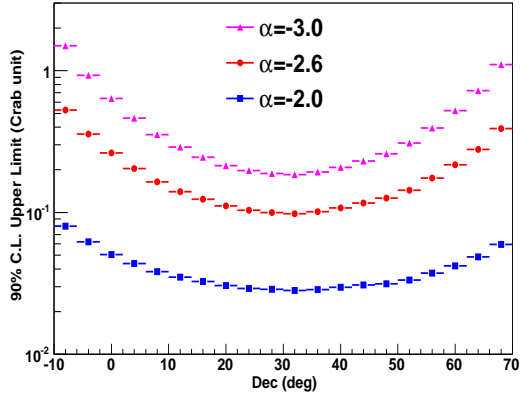}
  \caption{90\% C.L. upper limit on the integral flux above 0.1 TeV averaged on the right ascension direction, assuming a power-law spectrum. Different curves indicate sources with different spectral indices  -2.0, -2.6 and -3.0. The Crab unit is 9.90$\times$10$^{-10}$ cm$^{-2}$ s$^{-1}$ .}
  \label{fig5}
 \end{figure}

 It is interesting that Milagro discovered three extended sources around the Galactic plane \cite{abdo07a}, which have flux around 1 Crab Unit at 20 TeV. MGRO J1908+06 and MGRO J2031+41 have been observed by ARGO-YBJ with significance greater than 5 S.D., while there is no significant excess from the brightest source MGRO J2019+37. The upper limit derived using ARGO-YBJ data is lower than the flux reported by Milagro even after taking the source extension into account, which indicates that the flux of this source may  not be stable. Details on the analysis can be found in \cite{chensz11}.
\section{Conclusions}
The ARGO-YBJ experiment is an air shower array with large field of view
and has continuously monitored the northern sky since November 2007. Using data up to February 2011, we have presented the northern sky survey for VHE $\gamma$-ray point sources in the declination band between -10$^{\circ}$ and 70$^{\circ}$. There are four sources  observed with significance greater than 5 S.D.. The significance from Crab Nebula is about 17 S.D., which indicates that the cumulative sensitivity of ARGO-YBJ has reached 0.3 Crab units. The 90\% C.L. upper limits to the $\gamma$-ray flux from all the directions are obtained under the hypothesis of point sources with power-law spectra. The integral flux limits above 0.1 TeV vary from 0.09 to 0.53 Crab unit for Crab-like source depending on the declination, which is more restrictive than previous limits. Meanwhile, ARGO-YBJ is still collecting data as the only EAS array at TeV energies. The cumulative sensitivity will increase over time.

\section{acknowledgments}
 This work is supported in China by NSFC (No.10120130794),
the Chinese Ministry of Science and Technology, the
Chinese Academy of Sciences, the Key Laboratory of Particle
Astrophysics, CAS, and in Italy by the Istituto Nazionale di Fisica
Nucleare (INFN). We acknowledge the essential supports of W.Y. Chen, G. Yang,
X.F. Yuan, C.Y. Zhao, R. Assiro, B. Biondo, S. Bricola, F. Budano,
A. Corvaglia, B. D'Aquino, R. Esposito, A. Innocente, A. Mangano,
E. Pastori, C. Pinto, E. Reali, F. Taurino and A. Zerbini, in the
installation, debugging and maintenance of the detector.


\clearpage


\begin{thebibliography}{}
\bibitem{VHEweb}http://www.mppmu.mpg.de/~rwagner/sources/
\bibitem{albert08} Albert, J., et al. 2008, Science 320, 1752
\bibitem{abdo10}Abdo, A. A., et al. 2010, ApJS, 188, 405
\bibitem{hartman99} Hartman, R. C., et al. 1999, ApJS, 123, 79
\bibitem{aharon06} Aharonian, F., et al. 2006, ApJ, 636, 777
\bibitem{aharon02} Aharonian, F., et al. 2002, A\&A, 390, 39
\bibitem{atkins04} Atkins, R., et al. 2004, ApJ, 608, 680
\bibitem{amenom05} Amenomori, M., et al. 2005, ApJ 633, 1005
\bibitem{abdo07a}Abdo, A. A., et el. 2007, ApJ, 664, L91
\bibitem{aielli06} Aielli, G., et al. 2006, Nucl.
Instrum. Meth. A, 562, 92
\bibitem{aielli09b} Aielli, G., et al. 2009, Nucl.
Instrum. Meth. A, 608, 246
\bibitem{iuppa09}Iuppa, R., et al. 2009, ICRC, HE.1.1
\bibitem{he07} He, H.H.,  et al., 2007, Astropart. Phys., 27, 528
\bibitem{aielli09} Aielli, G., et al. 2009, Astrop. Phys.,
30, 287
\bibitem{barto11} Bartoli, B., et al. 2011, ApJ, 734, 110, or arXiv: 1106.0896v1
\bibitem{fleysher04} Fleysher, R., et al.
2004, ApJ, 603, 355
\bibitem{li83} Li, T.P.,\& Ma, Y.Q. 1983,
ApJ, 272, 317
\bibitem{abdo09c} Abdo, A.A., et al. 2009, ApJ, 700, L127
\bibitem{ameno10}Amenomori, M., et al. 2010, ApJ, 709, L6
\bibitem{helen83} Helene, O. 1983, Nucl. Instrum. Methods Phys. Res., 212, 319
\bibitem{chensz11} Chen, S.Z.,
 32$^{th}$ICRC, Beijing, OG.2.2, No.1005
\end{thebibliography}
\end{document}